\documentclass[pre,
showpacs, twocolumn
]{revtex4}

\usepackage[dvips]{graphicx}
\usepackage{amssymb,amsfonts,amsmath}
\usepackage{graphicx,color}
\usepackage{dcolumn}
\usepackage{epsfig}
\usepackage{latexsym}
\usepackage{color}
\usepackage{textcomp}

\newcommand{\Ln}{\mbox{$\left\langle L_n \right\rangle$}}

\newcommand{\Ree}{$\mathbf{R}_{ee}$}

\usepackage{graphicx,amsmath,amssymb,dcolumn,subfigure}
\usepackage{epsfig}

\begin{document}

\title{Direct visualization of flow-induced conformational transitions of single actin filaments in entangled solutions}
\author{Inka Kirchenbuechler,$^{1}$ Donald Guu,$^{2}$ Nicholas A. Kurniawan,$^{3}$ Gijsje H. Koenderink,$^{3}$
and M. Paul Lettinga,$^{2,4*}$}
\affiliation{$^1$ ICS-7, Forschungszentrum J\"{u}lich, D-52425 J\"{u}lich, Germany \\
$^2$ ICS-3, Forschungszentrum J\"{u}lich, D-52425 J\"{u}lich, Germany \\
$^3$ FOM Institute AMOLF, 1098XG Amsterdam, the Netherlands \\
$^4$ Department of Physics and Astronomy, Laboratory for Acoustics and Thermal Physics, KU Leuven,
Celestijnenlaan 200D, Leuven B-3001, Belgium}
\altaffiliation{Corresponding author: p.lettinga@fz-juelich.de}

\begin{abstract}

While semi-flexible polymers and fibers are an important class of material due to their rich mechanical properties, it remains unclear how these properties relate to the microscopic conformation of the polymers. Actin filaments constitute an ideal model polymer system due to their micron-sized length and relatively high stiffness that allow imaging at the single filament level. Here we study the effect of entanglements on the conformational dynamics of actin filaments in shear flow. We directly measure the full three-dimensional conformation of single actin filaments, using confocal microscopy in combination with a counter-rotating cone-plate shear cell. We show that initially entangled filaments form disentangled orientationally ordered hairpins, confined in the flow-vorticity plane. In addition, shear flow causes stretching and shear alignment of the hairpin tails, while the filament length distribution remains unchanged. These observations explain the strain-softening and shear-thinning behavior of entangled F-actin solutions, which aids the understanding of the flow behavior of complex fluids containing semi-flexible polymers.

\end{abstract}

\maketitle

Networks of semi-flexible polymers or fibers form the basis of many materials that we encounter in daily life. Concentrated dispersions of wormlike micelles are used to engineer the viscoelastic properties of industrial and consumer products \cite{Ezrahi2006}, while polysaccharides and self-assembled supra-molecular structures are used in tissue engineering \cite{Kim2008} and smart gels \cite{Hirst2008}. In biology, eukaryotic cells are mechanically supported by an internal cytoskeleton composed of protein filaments including  filamentous (F-)actin. These filaments can form striking non-equilibrium  patterns when they are subjected to cytoplasmic flows within plants \cite{Woodhouse2013} and animal embryos \cite{Ganguly2012} generated by molecular motor activity. The basis of understanding these phenomena is knowledge of the conformational response of the stiff polymers constituting the materials to an applied flow.

Here, we directly visualize the full three-dimensional (3D) contour of labeled \mbox{F-actin} subjected to shear flow, in order to resolve the microscopic basis of the macroscopic non-Newtonian flow response.
Our system lies in between the limiting cases of permanently cross-linked networks, where filaments do not relax, and of dilute non-interacting polymers, where filaments relax freely. Permanently cross-linked networks, which represent a model for cytoskeletal networks, show remarkable viscoelastic properties such as elasticity at low filament density and strong strain-stiffening behavior, where the stress increases with increasing strain \cite{Unterberger2013,Schmoller2010,Gardel2004}. Theoretical models can capture this behavior on the basis of microscopic properties of the filaments, such as the bending rigidity, length distribution, and cross-link density \cite{Broedersz2009}.

The rheology of \mbox{F-actin} solutions in the absence of cross-links \cite{Sato1985,Janmey1994,Koenderink2006,Semmrich2007,Semmrich2008} is comparatively poorly understood. The complex behavior of such solutions is due to entanglements between the filaments that form at concentrations above the overlap density, where diffusion takes place by reptation within tube-like confinement zones defined by the entanglements \cite{Kas94a}. The tube itself is not a rigid object, but describes rather a confining potential with a varying tube radius \cite{Glaser2010}. The linear viscoelastic response of such entangled \mbox{F-actin} solutions to small deformations that leave the tubes unaffected have been described by wormlike chain models \cite{Isambert96,morse98c}.
When starting up shear flow entangled \mbox{F-actin} solutions first display strain-stiffening, followed by strain-softening, where stress decreases with increasing strain \cite{Semmrich2007,Semmrich2008,Maruyama1974}. In addition, the viscosity of entangled F-actin decreases with increasing shear rate,  known as shear-thinning \cite{Kunita2012,Maruyama1974,Sato1985}.
These are key ingredients for the formation of flow instabilities, which often occur in concentrated semi-flexible polymers \cite{Wang2011} such as DNA \cite{Groisman00}, wormlike micelles \cite{Lerouge10} and also \mbox{F-actin} \cite{Kunita2012}. Though the connection between shear-thinning and flow instabilities is fairly well understood \cite{Olmsted99,Dhont99}, the microscopic mechanism of shear-thinning is still under debate mainly because real space information for nanoscopic systems like DNA and wormlike micelles is limited. \mbox{F-actin} is a particularly useful semi-flexible model polymer because the micron-sized lengths and relatively high stiffness allows imaging of the conformation at the single filament level by fluorescence microscopy \cite{Kas94a,Harasim2013,Kantsler2012,Steinhauser2012}.

2D filament tracking in microfluidic devices for dilute \mbox{F-actin} solutions, where the filaments do not interact, has shown that the conformational dynamics is dominated by the interplay between the Brownian motion of the filament ends and the shear flow \cite{Harasim2013}. Due to this competition, the filaments tumble in the gradient direction forming so-called hairpins, which contain highly curved segments between the stretched parts of the filaments. Here, the only relevant stress component is the shear stress, while reorientational motion takes place in the flow-gradient plane. The frequency of this tumbling motion in this direction was shown to decreases when the system is entangled \cite{Huber2014}. Since, however, stress can develop in all directions for entangled systems \cite{Moses94,Groisman00,Groisman98}, knowledge of the 3D contour of the filament is crucial for a full understanding of the mechanical response of these systems.

We use here concentrations just below and above the concentration where a confining tube can be defined \cite{Kas96a}. We achieve 3D in situ imaging by employing a counter-rotating cone-plate shear cell (Supplementary Fig.\,1) in combination with a fast confocal microscope \cite{Derks08}.
The advantage of this shear cell is that it induces a simple shear flow with a linear velocity gradient, while the presence of a zero-velocity plane guarantees that the filaments  stay sufficiently long in the field of view to obtain the full 3D contour.
This approach allows us to test predictions on the structural origin of strain-softening of entangled semi-flexible polymers \cite{morse99d,Fernandez2009}.
We quantitatively measure the effect of strain on the distribution of the local curvature. Moreover, a detailed analysis of the orientational distributions of the filaments reveals that entanglements are lost during strain deformation while hairpins are formed that are confined in the flow-vorticity plane. Thus we identify the mechanism for strain-softening and shear-thinning of entangled \mbox{F-actin} solutions.

\section{Results}
\subsection{Strain-softening and shear-thinning}
\begin{figure}[h]\centering
\includegraphics[width=0.45\textwidth]{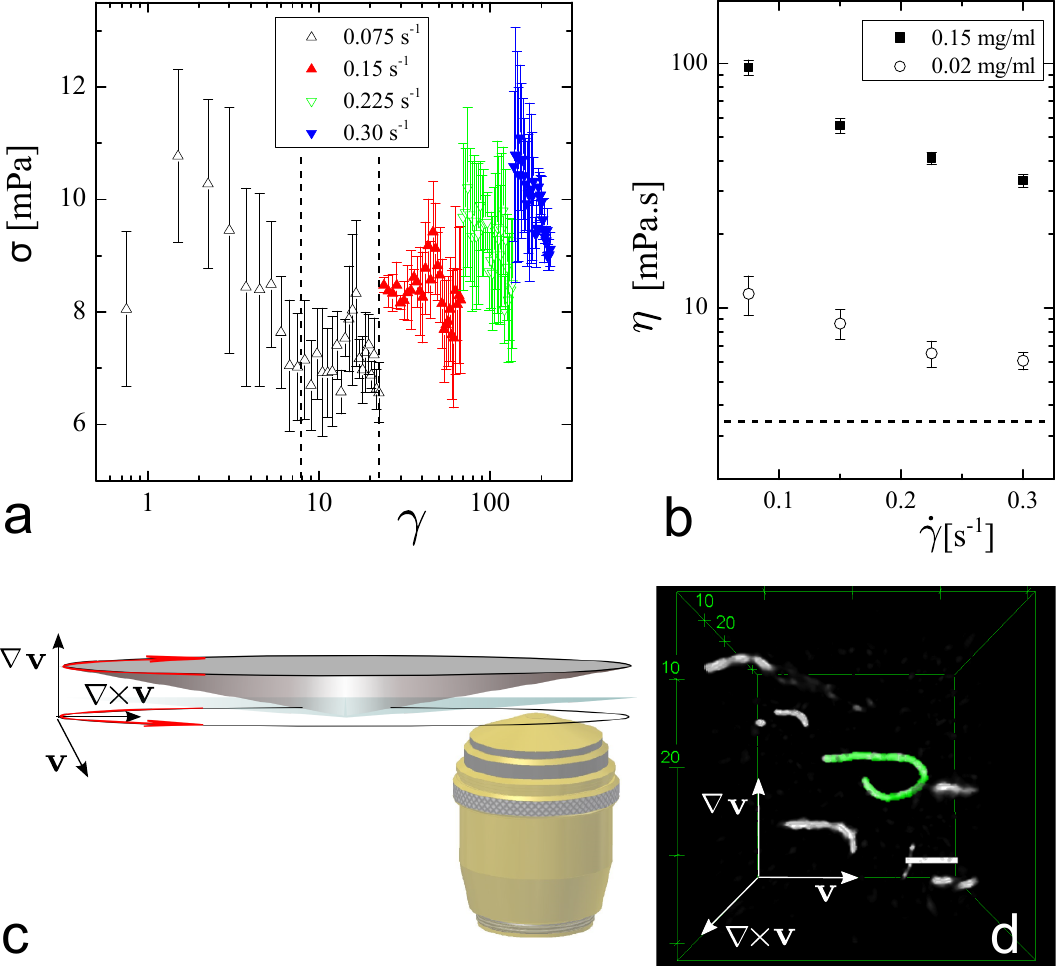}
\caption{\textbf{Rheology data and experimental setup} (a) Stress as a function of dimensionless strain where the shear rate is increased every five minutes at a concentration of $c_{high}$=\,0.15\,mg/ml. (b) Steady-state viscosity as obtained from the strain window where the stress is constant (see dashed lines in (a))  as a function of shear rate for the two concentrations. The dashed line gives the solvent viscosity. Error bars are due to the low torque.  (c) Schematic depiction of the counter-rotating cone-plate shear cell placed on an inverted microscope. (d) Typical reconstructed 3D stack of confocal images, showing a small fraction of labeled actin filaments embedded in a dark background of unlabeled \mbox{F-actin}. Green line: example of a tracked filament. Scale bar: 10\,\textmu m, tick unit: \textmu m. \label{fig_mm}}
\end{figure}

\begin{figure}[h]\centering
		\includegraphics[width=0.45\textwidth]{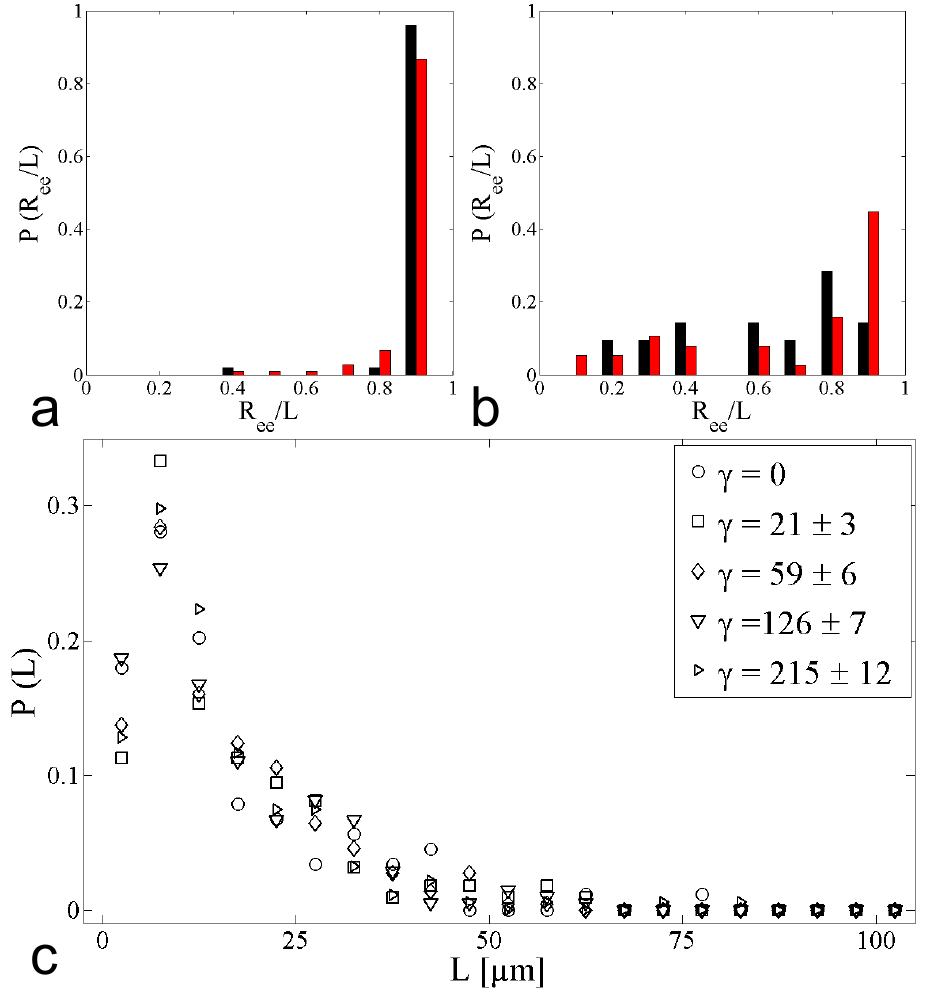}
		\caption{\textbf{Distribution of filament length parameters} (a) Ratio of end-to-end distance \Ree\;over the filament length L\label{fig_RL} for short filaments $L<13$\,\textmu m and (b) long filaments $L>21$\,\textmu m. Black bars: zero strain and red bars strain $\gamma=215\pm12$. (c) Filament length distribution at different dimensionless strain values. All distributions were determined for $c_{high}$. \label{RL_length} }
\end{figure}

We subjected \mbox{F-actin} solutions of 0.02 and 0.15\,mg/ml to shear flow , increasing the shear rate from 0.075 to $0.3\,s^{-1}$ in 4 steps of 5 minutes, while keeping track of the total acquired strain, see section \ref{sec_methods}. The stress as a function of strain at 0.15\,mg/ml is plotted in Fig.\,\ref{fig_mm}a. The stress decreases with increasing strain after start-up of the shear flow at the lowest shear rate. This behavior is indicative for strain-softening. After about 8 strain units the stress stays constant and the viscosity of the system can be determined. In Fig.\,\ref{fig_mm}b the viscosity is plotted as a function of shear rate for both concentrations. Both systems display a clear shear-thinning behavior. Thus we observe both strain-softening and shear-thinning. The viscosity at 0.02\,mg/ml is an order of magnitude less than that at 0.15\,mg/ml and approaches for high shear rates the buffer viscosity, exemplifying the effect of entanglement. From now on we refer to these samples as $c_{high}$=\,0.15\,mg/ml and $c_{low}$=\,0.02\,mg/ml, where the indices refer to the viscosity of the samples.

\subsection{Global and local information as obtained from imaging}

To visualize shear-induced conformational transitions of entangled individual actin filaments, we embedded trace amounts of fluorescently labeled actin filaments in the host dispersion. For imaging we used a home-built counter-rotating cone-plate cell mounted on an inverted microscope  (Fig.\,\ref{fig_mm}c) that was equipped with a multi-pinhole confocal  scanning  head,  allowing  frame rates of 8 frames  per second. We obtain the full 3D contour of on average 150 filaments per analyzed dimensionless strain.
Fig.\,\ref{fig_mm}d shows several filaments, including one for which the contour is fitted. The coordinate system of this fitted contour, parametrized by $\mathbf{r}_j$ ($j$ is the index of the coordinates along the contour), is given by the velocity direction (along the shear direction), gradient direction (along the gap direction) and vorticity direction.
From $\mathbf{r}_j$  we can extract information about the global filament conformation, such as the filament contour length $L$ and its end-to-end vector \Ree.
Fig.\,\ref{RL_length}a and b show the distribution of the ratio \Ree$/L$ for $c_{high}$, which is a measure for the degree of stretching.
This distribution does not show a significant change with strain for the short filaments ($<13$\,\textmu m, Fig. \ref{RL_length}a).
For the long filaments ($>21$\textmu m, \ref{RL_length}b), the distribution widens towards small \Ree$/L$ while a pronounced peak is formed for \Ree$/L \rightarrow 1$. This means that the filaments become more stretched and at the same time more bent, which is indicative for the formation of hairpins.
Fig.\,\ref{RL_length}c displays the length distribution at different dimensionless strain values for $c_{high}$. The distribution does not shift to lower values of filament lengths with increasing strain, so that we conclude that shear-softening and shear-thinning is not due to rupture of actin filaments.

We will characterize the filament conformations by extracting the local curvature, $\kappa_j$, tangent vector, $\hat{T_j}$, and binormal vector, $\hat{B_j}$, along the contour, see Fig.\,\ref{fig_examples}a. To calculate these values we average over two neighboring points in the contour. Thus, the contour is split up in segments of an average length of \Ln=\,2.36\,\textmu m (Supplementary Note\,1).
Fig.\,\ref{fig_examples}b and c display the projections of two typical examples of filaments strained at $c_{low}$ (b) and $c_{high}$ (c).
At $c_{low}$, which corresponds to the onset of entanglements, we find filaments with highly curved segments (region I) and segments that are stretched and orientated with the flow direction (region II). At $c_{high}$, the filament has a hairpin conformation (region III), with two dangling ends which are stretched in the flow direction (region IV). Strikingly, for $c_{high}$ the hairpin is confined to the flow/vorticity plane, with its binormal vectors (blue arrows) pointing in the gradient direction Fig.\,\ref{fig_examples}c. This behavior is markedly different from the behavior observed for filaments in dilute solutions \cite{Harasim2013}, which tumble in the gradient direction. The behavior at $c_{low}$ is intermediate.

\subsection{Filament stretching and bending in shear flow}

\begin{figure}[h]\centering
\includegraphics[width=0.45\textwidth]{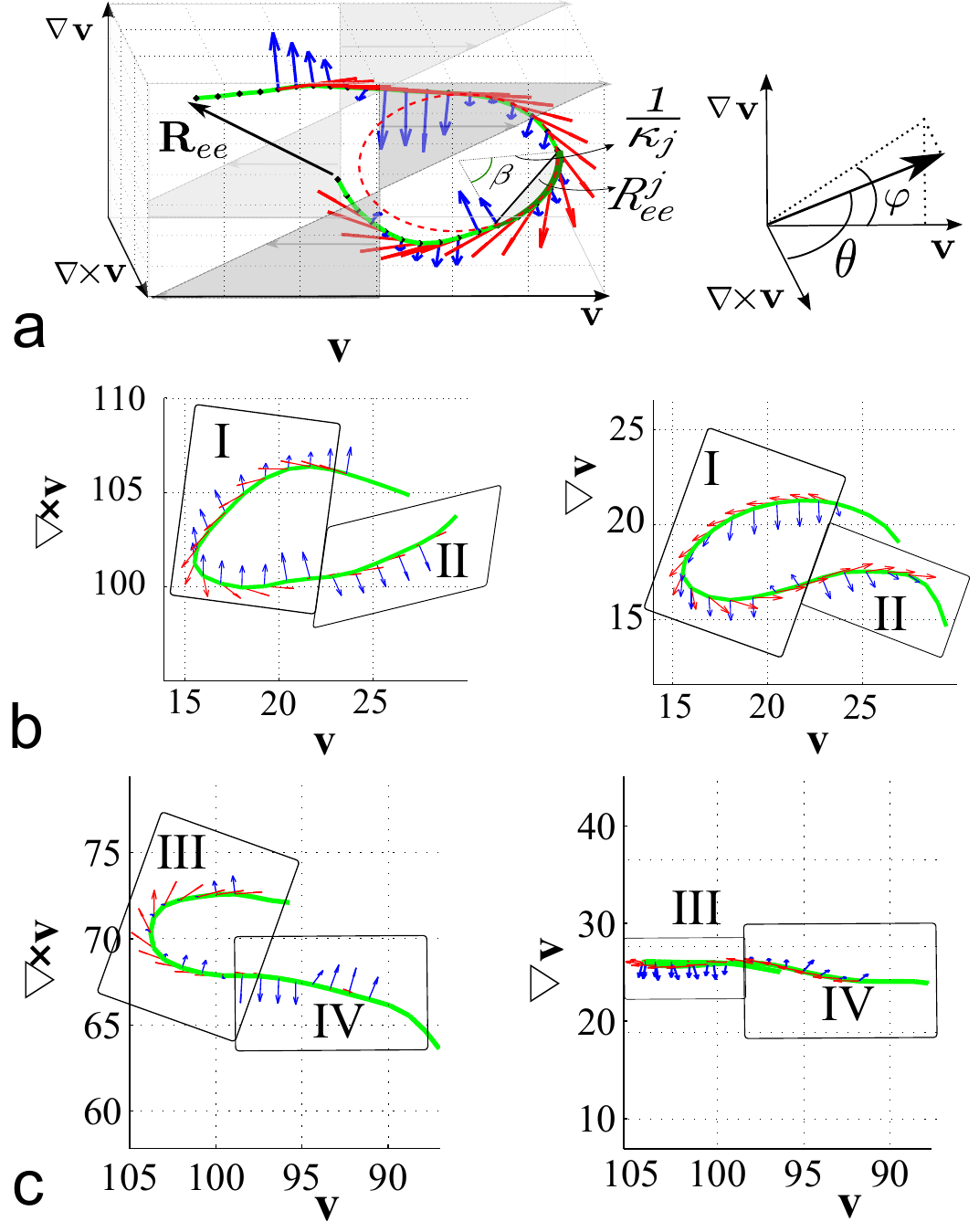}
\caption{\textbf{Tracked conformations of actin filaments} (a) Example of a tracked filament (Fig.\,\ref{fig_mm}d) with its end-to-end vector, \Ree\;(black). The filament is subdivided in segments $j$, allowing us to determine the local orientation in terms of tangent $T_j$ (red) and binormal $B_j$ (blue) vectors, and the local curvature using the local end-to-end vector \Ree$^j$. Part of the hairpin can be described a circle with curvature $\kappa_j$. The relevant angles characterizing the orientation are indicated on the right. (b,c) Typical conformations of actin filaments  at $c_{low}$ (b) and $c_{high}$ (c). Left: views in the flow/vorticity plane. Right: corresponding views in the flow/gradient plane.
At $c_{low}$, the filament has a highly curved segment (region I), and stretched segments that are somewhat oriented in the flow direction (region II). It shows no confinement in any plane. For $c_{high}$, the filament also has a highly curved segment (III) and a stretched segment (IV) and is confined to the flow/vorticity plane. Tick unit: \textmu m.\label{fig_examples}}
\end{figure}

The typical configuration of a sheared filament as displayed in Fig.\,\ref{fig_examples}a shows a hairpin, which is characterized by two stretched tails connected by a bent part. The formation of these hairpins suggests that the distribution of the curvature of the segments changes when entangled semi-flexible polymers are sheared. Stretching and bending of filaments both contribute to the free energy of a filament. Usually this is connected to \Ree\; which measures the stretching of a filament \cite{Winkler03}. \Ree, however,  can be small for hairpins, which can thus be incorrectly taken to suggest that the entropic contribution to the energy is low. This is clearly a flaw because the main part of the filament is stretched. One therefore needs to analyze the filaments at smaller length scales. We do this by using the local curvature $\kappa_j$, which is related to the scaled end-to-end vector $x_j=R_j/\Ln$ (Fig.\,\ref{fig_examples}a). In this analysis we only used filaments with length $L>21$\,\textmu m to assure good statistics and sufficient flexibility ($L/l_p > 1$). On average we analyzed about 1200 segments per image stack, allowing us to obtain distributions for the curvature $P(\kappa_j)$.

\begin{figure}[h]\centering
\includegraphics[width=0.45\textwidth]{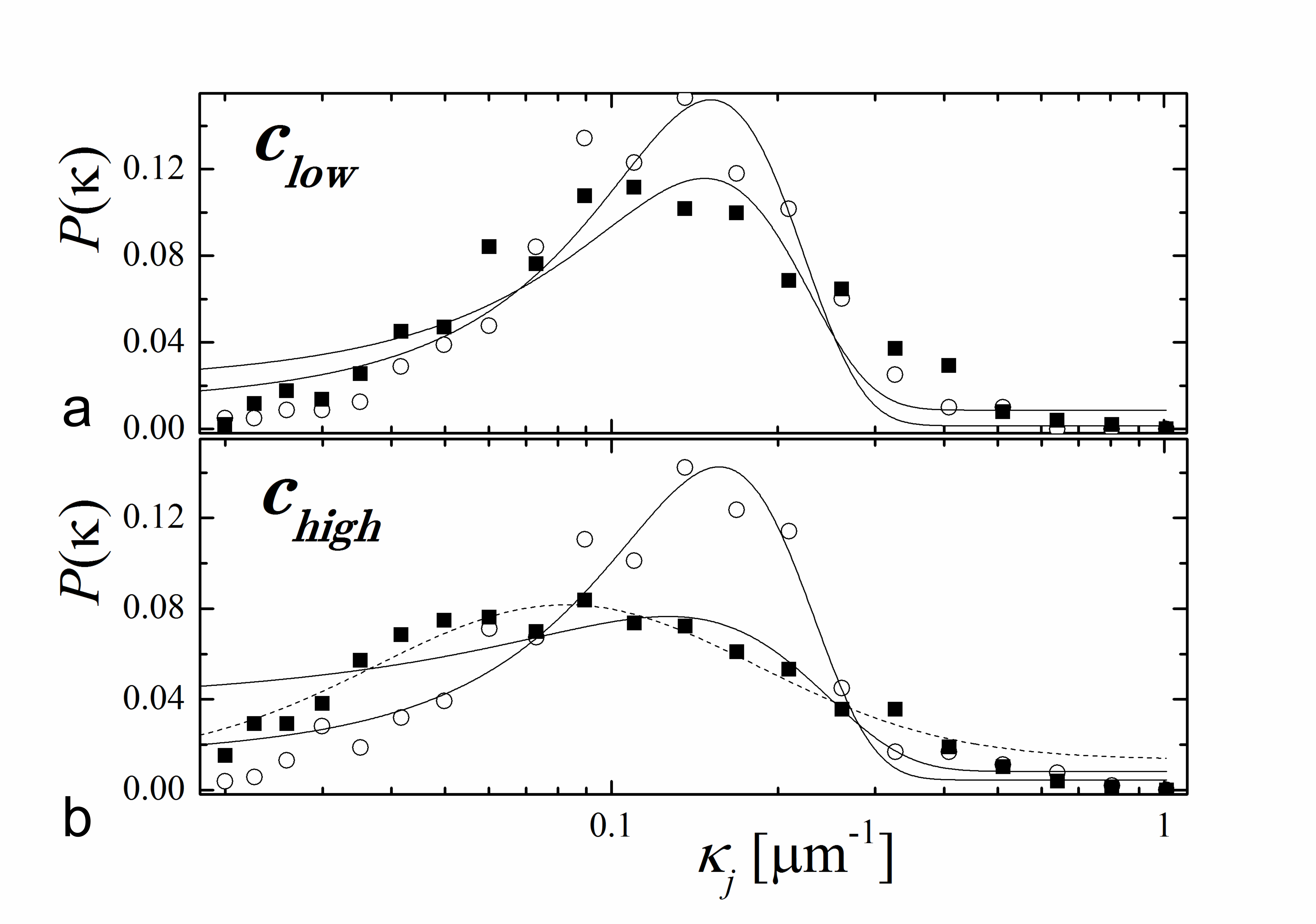}
\caption{ \textbf{Distribution functions of $\kappa_j$} (a) $c_{low}$ (b) $c_{high}$. Open symbols: zero strain $\gamma= 0$. Solid symbols: high strain, $\gamma= 215\pm12$. Dashed line is fit to a log-norm distribution, while the solid lines are fits to a Gaussian distribution.\label{Fig_DisKappa}}
\end{figure}

The distribution of $\kappa_j$ are plotted in Fig.\,\ref{Fig_DisKappa} for zero shear strain and high strain ($\gamma= 215\pm12.$). For both concentrations the equilibrium curvature distribution is well-described by a Gaussian distribution. The distribution in equilibrium deviates from earlier studies of unsheared F-actin \cite{Romanowska2009}, probably due to the different sample environment and relatively high concentrations of filaments that are used in this study. For the low concentration the distribution remains unchanged when straining the system. In contrast, at the high concentration we observe that the distribution takes the form of a log-normal distribution. Compared to the unsheared situation, the peak of the distribution shifts towards smaller values of $\kappa_j$, showing that the majority of segments are more stretched. In addition the distribution also has a long tail with a power law form, showing that strain induces curvature in the system. These observations hint that at $c_{high}$  energy is stored in the system, but not at $c_{low}$.
This is consistent with the presence of entanglements at $c_{high}$. However, these observations do not yet explain the strain-softening behavior.

\subsection{Orientation of shear-induced hairpins}

\begin{figure*}[t]\centering
\includegraphics[width=0.95\textwidth]{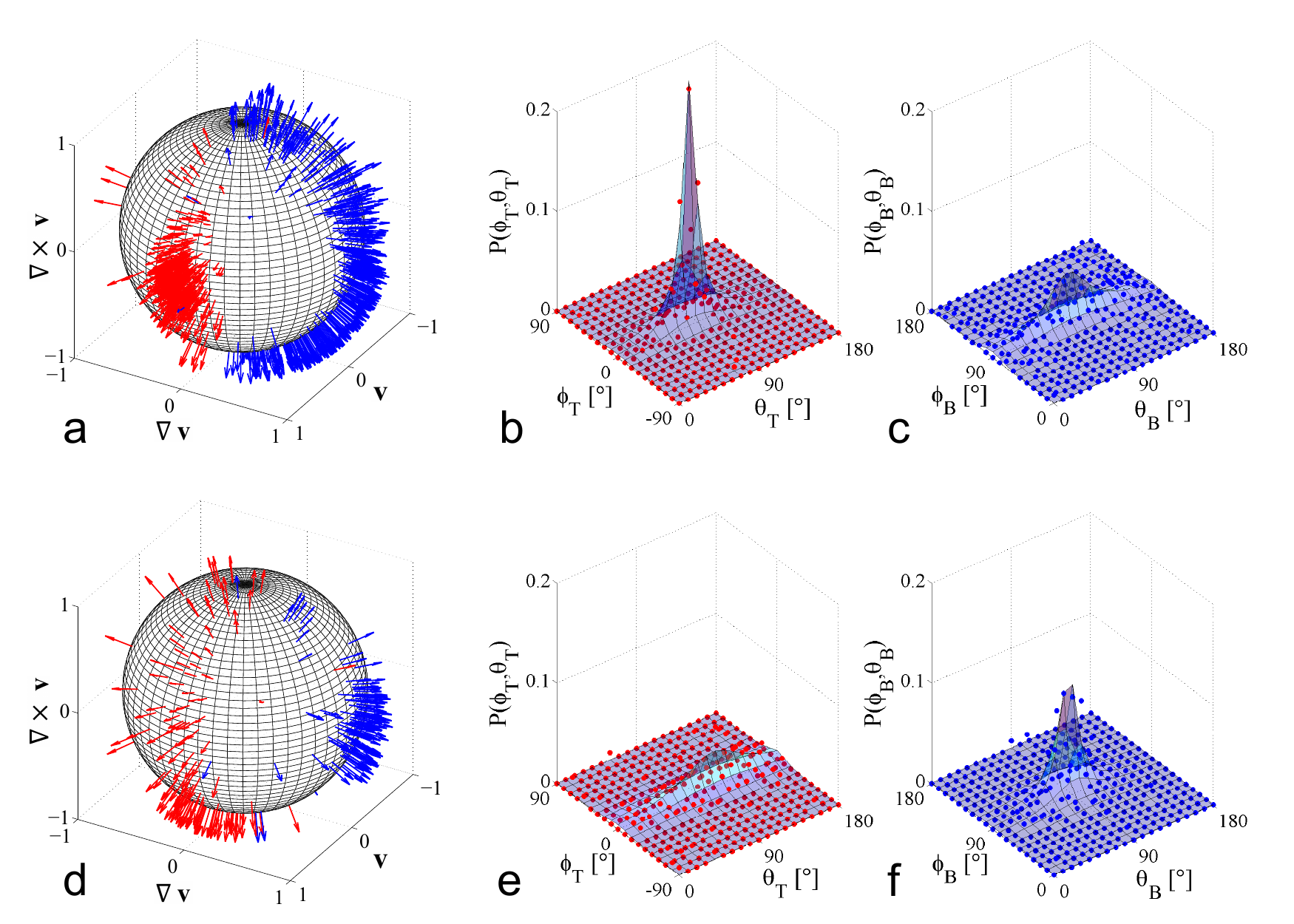}
\caption{\textbf{Distribution of orientational order parameters} (a,d) Tangent (red) and binormal (blue) vectors on a unit sphere. Corresponding angular distribution as given by their angles $\theta$ and $\phi$ with a 2D Lorenzian for tangent (b,e) and binormal distribution (c,f). We used $c_{high}$=\,0.15\,mg/ml, filament length $>21$\,\textmu m, $\gamma=215\pm12$. The top row are results for stretched segments (cut-off curvature $\kappa_j<0.1$\,\textmu m$^{-1}$) and bottom row are results for the bent segments ($\kappa_j>0.2$\,\textmu m$^{-1}$).\label{fig_distribute}}
\end{figure*}

To quantify the orientations of the filaments, we collected all local tangent and binormal vectors along the filament contour, as plotted in Fig.\,\ref{fig_distribute}a,d on a unit sphere, and calculated the corresponding two-dimensional angular orientational distribution functions which are best fitted with a 2D Lorentzian given by
$f(\theta,\phi)=a/\left((\frac{\theta-\Delta\theta}{w_{\theta}})^2+(\frac{\phi-\Delta\phi}{w_{\phi}})^2+1\right)$. Here the angles $\varphi$ and $\theta$, which set up the vectors $\hat{T_j}$ and $\hat{B_j}$, are defined as shown in Fig.\,\ref{fig_examples}a right. As examples we show in  Fig.\,\ref{fig_distribute} the distribution for $c_{high}$ after straining the sample to $\gamma=$\,215$\pm$12, separating the orientation of the stretched segments, using all segments with $\kappa_j<0.1$\,\textmu m$^{-1}$ (top) and bent segments using all segments with $\kappa_j>0.2$\,\textmu m$^{-1}$ (bottom).

The stretched segments display a strong shear-alignment of the tangent: Fig.\,\ref{fig_distribute}a,b shows that the segments all point in the flow direction. Note however that the distributions are  slightly biaxial, i.e. not symmetric around the maximum. The binormal distribution has a low degree of order and is strongly biaxial, Fig.\,\ref{fig_distribute}c. This can be rationalized as follows. When the tangent is well aligned, then there is no clearly defined plane that is spanned by two sequential segments. Thus the binormal is ill-defined though per definition in the plane perpendicular to the tangent: the distribution in Fig.\,\ref{fig_distribute}c is very sharp with respect to $\phi_B$ and very broad with respect to $\theta_B$. On the contrary, the bent segments have a pronounced uniaxial ordering of the binormal pointing in the gradient direction (Fig.\,\ref{fig_distribute}d,f), while the distribution of the tangent is biaxial (Fig.\,\ref{fig_distribute}e), which is to be expected for hairpins, see for example region I in Fig.\,\ref{fig_examples}b.

In order to quantify separately the orientational behavior of the stretched and bent parts of the filaments, we use the distribution functions $f(\theta,\phi)$ to calculate the orientational order tensors $\bar{S}_{T}=\int^{\pi}_{0}\int^{2\pi}_{0}d\theta d\phi\sin(\phi) f(\theta_{T},\phi_{T}) \hat{T}\hat{T}$ and similarly $\bar{S}_{B}$. For our purpose the traceless diagonalized form $\bar{Q}=\frac{1}{2}(3\bar{S}-\mathbf{I})$ is particularly useful since Fig.\,\ref{fig_examples} and Fig.\,\ref{fig_distribute}b both suggest that the orientational distributions can be biaxial. $\bar{Q}$ may be written as

\begin{eqnarray}\label{Eq_Qtensor}
\bar{Q}_{T,B}\:=\:\left(  \begin{array}{ccc}
                -\frac{1}{2}\lambda_{T,B}-\eta_{T,B}\; & \;0 \; & \;0 \\
                0\; & \;-\frac{1}{2}\lambda_{T,B}+\eta_{T,B}\; & \;0 \\
                0\; & \;0\; & \; \lambda_{T,B}
                                      \end{array} \right) \:,
\end{eqnarray}

\noindent where $\lambda_{T,B}$ is the orientational order parameter of the main orientation axis and  $\eta_{T,B}$ parametrizes the biaxiality of the system. 

\begin{figure}[h!]\centering
\includegraphics[width=0.5\textwidth]{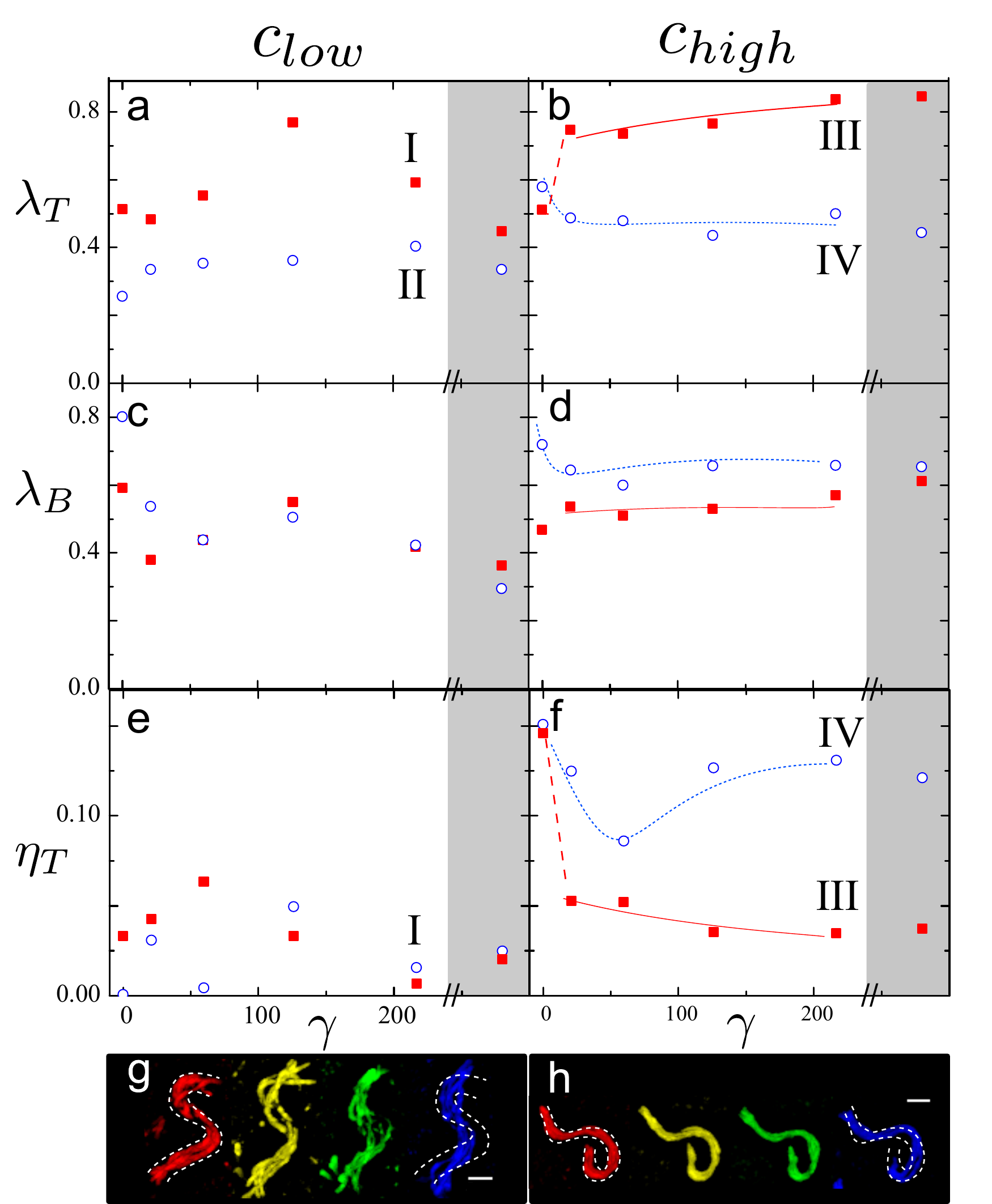}
\caption{\textbf{Orientational order parameters} (a,b) Tangent and (c,d) binormal vector. (e,f) Degree of biaxiality. The solid symbols are the results for stretched segments ($\kappa_j<0.1$\,\textmu m$^{-1}$) and the open symbols are the results for the bent segments ($\kappa_j>0.2$\,\textmu m$^{-1}$). The grey zone indicates data points about 150\,s after cessation of shear flow. Also shown is the overlay of the 2D projection of filaments  (g and h) in four sequential time windows following cessation of shear flow (time average $45$\,s each). The dotted white lines indicate the confining tube for the first 45\,s following cessation of shear flow. Scale bar 5\,\textmu m. The left column is data for $c_{low}$ and the right column for $c_{high}$. The lines in (b,d,f) are guides to the eye. I-IV refer to the regions shown in Figure \ref{fig_examples}.  \label{fig_orient}}
\end{figure}

\begin{figure}[h!]\centering
		\includegraphics[width=0.45\textwidth]{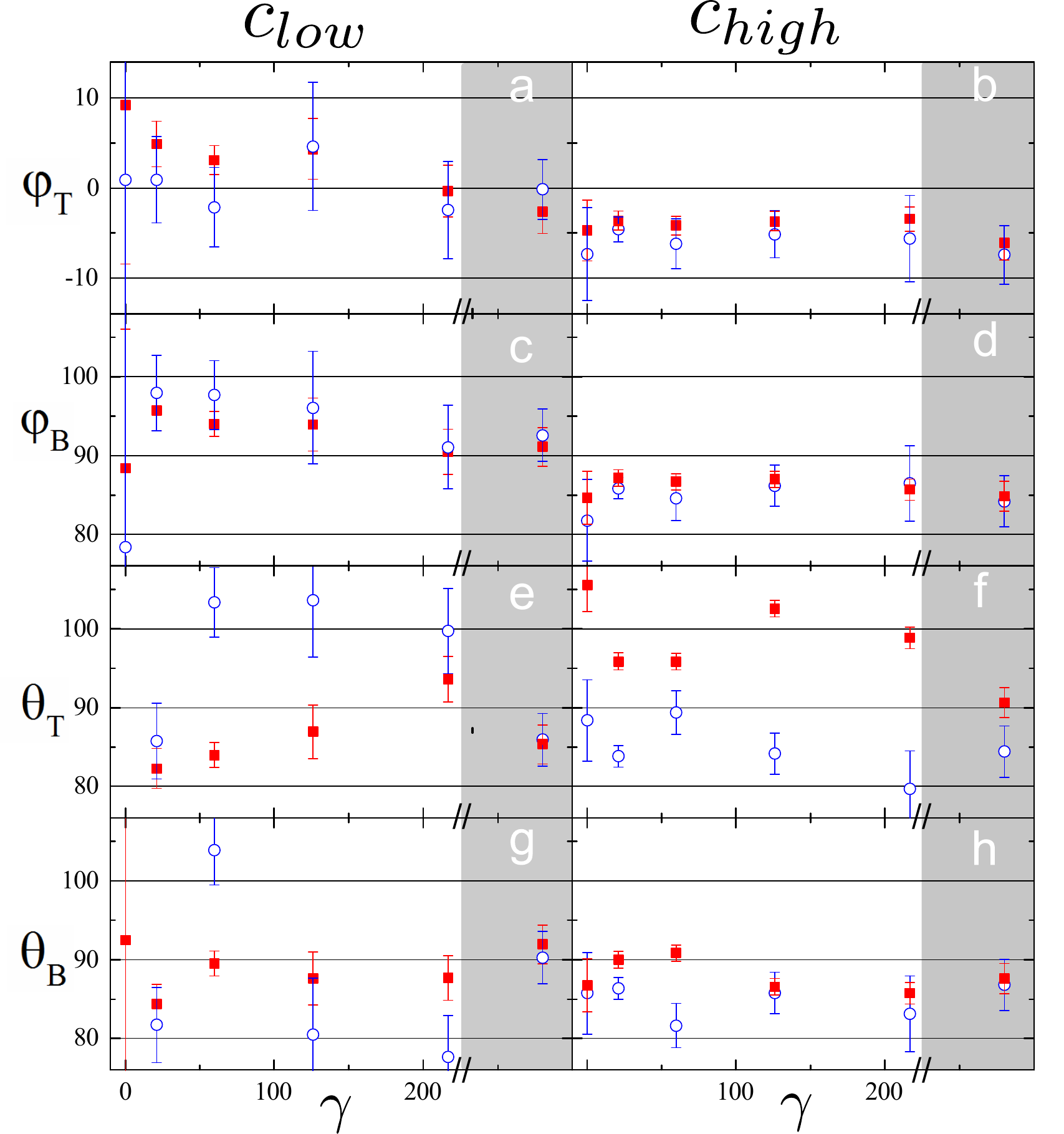}
		\caption{\textbf{Angles of the eigenvectors} Angles correspond to the highest eigenvalue of the tangent tensor (a,b,e,f) and the binormal tensor (c,d,g,h). The solid symbols are the results for stretched filaments ($\kappa_j<0.1$\,\textmu m$^{-1}$) and the open symbols are the results for the bent segments ($\kappa_j>0.2$\,\textmu m$^{-1}$). The grey zone indicates data points 150\,s after cessation of shear flow.
\label{fig_angles} }
\end{figure}

We will now discuss the behavior of these parameters for the four features that we indicated in Fig.\,\ref{fig_examples}b and c: segments with high curvatures of $\kappa_j>0.2$\,\textmu m$^{-1}$ for $c_{low}$ (I) and $c_{high}$ (III), and segments with low curvatures of $\kappa_j<0.1$\,\textmu m$^{-1}$ for $c_{low}$ (II) and $c_{high}$ (IV).
The stretched segments at $c_{high}$ (IV in Fig.\,\ref{fig_examples}) clearly display a high degree of ordering of the tangent with $\lambda_T\rightarrow 1$ (solid symbols in Fig.\,\ref{fig_orient}b). The ordering is uni-axial (solid symbols in Fig.\,\ref{fig_orient}f) and almost along the flow direction, as expected (solid symbols in Fig.\,\ref{fig_angles}c and f).
$\lambda_T$ is higher for $c_{high}$ than for $c_{low}$ (compare Fig.\,\ref{fig_orient}a and b, II and IV), which shows that entanglements enhance ordering when the sample is sheared.
Interestingly, $\lambda_T$ jumps immediately to a high value when strain is applied for $c_{high}$. On the contrary, $\lambda_T$ displays a moderate increase with increasing strain for $c_{low}$, while the corresponding eigenvector tilts towards the flow direction (solid symbols in Fig.\,\ref{fig_angles}a and e). This complies with measurements on wormlike micelles \cite{Helgeson2009}.
Thus, the entanglements not only cause stretching of parts of the filaments, as seen from Fig.\,\ref{Fig_DisKappa}, but also strong shear-alignment of these stretched segments along the flow direction.

Whilst this shear-alignment is a well known phenomenon, the orientational behavior of segments with high curvature (I and III in Fig.\,\ref{fig_examples}) is \textit{a priori} not obvious. The binormal is a particularly informative parameter since it is oriented along the normal of the plane spanned by a curved segment. When there is a train of segments with similar curvature, which is the case in the bent part of a hairpin, then binormal vectors point in the same direction. This can be seen for $c_{high}$, where the blue arrows in region III of Fig.\,\ref{fig_examples}c on the right all point in the gradient direction. The behavior of the binormal order parameter $\lambda_B$ confirms this observation. Indeed $\lambda_B$ is higher for the highly curved than for the stretched segments; compare open and solid symbols in Fig.\,\ref{fig_orient}d.
The eigenvectors belonging to $\lambda_B$ point along the gradient direction (Fig.\,\ref{fig_distribute}f and open symbols in Fig. \,\ref{fig_angles}d and h) and therefore also the plane spanned by the highly curved segments.
The fact that the binormal is well defined for $\kappa_j>0.2$\,\textmu m$^{-1}$ implies that the ordering of the tangent is low and biaxial. This is indeed confirmed by Fig.\,\ref{fig_orient}f, where the biaxiality of the tangent $\eta_T$ is plotted: for $c_{high}$ the biaxiality of the tangent $\eta_T$ increases with strain when $\kappa>0.2$\,\textmu m$^{-1}$.
We therefore conclude that the plane spanned by the bent part of the hairpin, which we labeled III in Fig.\,\ref{fig_examples}c, indeed is located in the flow-vorticity plane, with its normal pointing in the gradient direction.

The behavior of the binormal and biaxiality is distinctively different for $c_{low}$. No increase in $\lambda_B$ nor in $\eta_T$ is observed with increasing strain (Fig.\,\ref{fig_orient}c,e). Thus, there is no well defined orientation of hairpins at the low concentration, as is exemplified in region I of Fig.\,\ref{fig_examples}b.
In Fig.\,\ref{fig_orient} we also plot the points measured about 150\,s after cessation of shear flow. We note a clear difference between the relaxation times at low and high concentration: for $c_{low}$ the orientation is partly lost, whereas for $c_{high}$ the orientations remain unchanged. This indicates that the relaxation time at $c_{high}$ is indeed much longer than for $c_{low}$. A time series after cessation of flow confirms these observations: the shape is mainly lost for $c_{low}$ (Fig.\,\ref{fig_orient}g), while at $c_{high}$ the filament conserves its shape (Fig.\,\ref{fig_orient}h).

\section{Discussion}
By directly visualizing the full 3D conformation of individual actin filaments within entangled \mbox{F-actin} solutions, we can show how entanglements influence the conformation of the filaments in response to shear flow.

First, the distribution of curvature for $c_{low}$, at the onset of the entangled regime, remains unchanged when applying shear flow, while the number of stretched as well as bent segments increases for $c_{high}$  when applying shear, see Fig. \ref{Fig_DisKappa}. This explains the much higher viscosity for $c_{high}$. Second, we observe at $c_{high}$ that hairpins form in shear flow, which tilt into the flow-vorticity plane. The behavior of entangled filaments is in marked contrast with the dynamics of filaments in dilute solutions which tumble in the gradient direction \cite{Harasim2013}. Our observations also explain why tumbling was previously shown to be strongly reduced at concentrations three times higher than $c_{high}$ \cite{Huber2014}, although the shear rates used in this reference were $\mathcal{O}(10^2)$ higher. We find that the response at $c_{low}$ lies between these two extremes, since there is no well defined plane into which the filaments turn. In this case we find shear-thinning which is purely due to increased alignment of the filaments, as we conclude from Fig.\,\ref{fig_orient}a and e.

We will now try to relate the formation of the strongly aligned hairpins with the strain-softening in entangled solutions. Hairpins are generally viewed as a signature of an entanglement. Two entangled filaments will strongly bend around the point where they are entangled when they are moved in opposite directions faster than the time they need to relax. Mechanically this leads to strain-stiffening, while pairs of hairpins with the orientations of the bent segments roughly perpendicular to each other result in a very flat distribution of the binormal of the bent segments. We observe, however, exactly the opposite:  the binormal of the bent segments is highly aligned when straining the system at $c_{high}$. Moreover, there is a strong increase in the number of stretched segments and a marginal increase in the number of bent segments and we predominantly find only one hairpin per filament, see Fig. \ref{Fig_DisKappa}b. These observations strongly hint that entanglements disappear as the system is strained.
Since the hairpins and their stretched tails are strictly located in the flow-vorticity plane, they have no components in the gradient direction that cause the shear stress, which results in strain softening. The filaments slide over each other facilitating lamellar flow. These findings are contrary to theoretical predictions that hairpins cause strain-stiffening. However, the theory considered the contour lengths to be  much longer than the persistence length \cite{morse99d,Fernandez2009}, while in our experiments they are of the same order. Theory does predict an instability where shear deformation pushes out contacts between filaments causing strain-softening \cite{Fernandez2009}. This strain-induced loss of entanglements, very similar to the convective constraint release \cite{Marrucci1996}, is exactly what we find.

There are no predictions of the shape of the filaments at high strains after strain softening. We find experimentally that the effect of the surrounding filaments is to confine the hairpins in the flow-vorticity plane. Instead of a confining tube as can be defined in equilibrium, there are now confining planes, without entanglements. This can also be appreciated from the movie (Supplementary Movie\,1), where fluctuations in the vorticity directions are observed, but not in the gradient direction, in contrast to the movie of a filament at $c_{low}$ (Supplementary Movie\,2).
While the hairpins immediately form and can be related to the strain-softening, we also observe shear thinning, see Fig. \ref{fig_mm}b. This behavior is likely to relate to the small but significant increase in the orientational ordering of the stretched segments (solid symbols in Fig \ref{fig_orient}b) and decrease in the biaxiality (solid symbols in Fig \ref{fig_orient}f) parameterizing the flow alignment of the segments.
These new scenarios for strain-softening and shear-thinning exclude the need for scission of the \mbox{F-actin} filaments as a pathway to explain non-linear rheology, which is known for living polymers such as wormlike micelles \cite{Lerouge10}. Indeed, we observe in Fig. \ref{fig_RL}b no change in the filament length distribution over the full measured range of strain.

The mechanism for stress release by the formation of disentangled sliding hairpins could be a precursor for the formation of flow instabilities, which is often related with local reorganization of the constituting particles \cite{Lerouge10}. The distinct orientation of the hairpins also suggests that a normal stress builds up which could lead to flow instabilities.
Polymer solutions \cite{Moses94,Groisman00,Groisman98}, polymer melts \cite{Oda1978} and sticky carbon nanotubes \cite{Lingibson04} are all systems that display pronounced normal stresses as well as flow instabilities. Flow instabilities have also been observed for actin dispersions at higher concentrations than we used \cite{Kunita2012}.
We did not find any signature of such a behavior, scanning a significant part of the gap of the shear cell at a fixed position from the center of the shear cell. This could be due to the limited strain applied to the system as well as the relatively low filament concentration as compared to ref. \cite{Kunita2012}.

In conclusion, we believe that the mechanism of stress release we identified here may be generally valid for solutions of semi-flexible polymers, including supramolecular systems \cite{vanderGucht03b,Kouwer2013,lonetti08,Lerouge10}. Thus, our findings will aid to the understanding of the complex flow behavior of such systems. Likewise, this mechanism could impact the self-organization of cytoskeletal filaments in response to intracellular shear flows created by processes like cytoplasmic streaming \cite{Woodhouse2013,Ganguly2012}.

\section{Methods}\label{sec_methods}
\subsection{Protein purification and sample preparation}\label{ssec_matSample}
G-actin was isolated from rabbit skeletal muscle \cite{Spudich1971}, stored in G-buffer solution (5.0\,mM Tris, 0.2\,mM CaCl$_2$, 0.2\,mM ATP, 0.2\,mM DTT, pH\,7.4, 21$^\circ$C) at 4$^\circ$C. Fluorescently labeled filaments were obtained by polymerizing G-actin (0.2\,mg/ml) for 1\,h at 21$^\circ$C by adding 10 vol\% 10x\,F-buffer solution (20\,mM Tris, 2.0\,mM CaCl$_2$, 1.0\,M KCl, 20\,mM MgCl, pH\,8.5, 21$^\circ$C, 5.0\,mM ATP, 2.0\,mM DTT) in the presence of an equimolar amount of Atto647N-Phalloidin (Atto-Tec). Unlabeled G-actin (0.2-0.3\,mg/ml) was similarly polymerized, but in the presence of unlabeled phalloidin (Sigma/P2141). These filaments have a persistence length of $l_p$ 9-18\,\textmu m \cite{Ott93,Git93,Isa96,Mam09} Samples were prepared by diluting labeled filaments in a 1:2000 ratio with GFS-buffer solution (10\% 10x\,F-buffer, 60\% sucrose in G-buffer) and mixing this solution in equal volume with unlabeled \mbox{F-actin} to reach final concentrations of 0.02 and 0.15\,mg/ml. The final buffer solution thus contained 30\% sucrose, which reduced the off-rate of labeled phalloidin \cite{DeLaCruz1994}, improving the signal to noise ratio during image acquisition. All measurements were done at 21$^\circ$C.

\subsection{Shear cell and Microscopy }\label{ssec_matSetup}
To produce shear flows with a well-defined linear velocity gradient, we used an adapted version of the counter-rotating cone-plane shear cell used in ref.\,\cite{Derks04} (Supplementary Fig.\,1). It consists of a bottom glass plate (diameter 80\,mm, thickness 170\,\textmu m, Menzel) which is fixed by two Teflon rings that are pressed together between a titanium plate at the bottom and a stainless steel block at the top. In this block there is a hole where the top cone is inserted, which is also made out of stainless steel. Both the glass plate and steel cone can move independently and are in our experiments moved counter clockwise. We used an Epiplan-Neofluar 50x/1.0 Pol objective (Zeiss) . The shear cell was mounted on an inverted microscope (Zeiss/Axiovert 200 M), equipped with a multi-pinhole-confocal system (VisiTech/VT-Infinity-I) and an Epiplan-Neofluar 50x/1.0 Oil Pol Objektiv (Zeiss). An Argon/Krypton laser (Spectra Physics/Stabilite 2250) operating at 647\,nm was used for excitation of the fluorescent dye.
An observation area of 151\,\textmu m x 151\,\textmu m was imaged onto an EMCCD-camera (Andor/iXon DU-897) operated with IQ software. Confocal stacks consisted of 51 frames taken at 1\,\textmu m intervals at a rate of 7.2\,s per stack. The difference in the geometrical and the optical path lenght, due to the different refractive indices of the immersion oil ($n_{oil}=1.518$) objective and the actin solution ($\approx n_{water}=1.33$) was determined by two independent methods. First we measured the refractive indices of the immersion oil and the actin solution and calculated the correction factor $n_{k}$ based on the ratio of those indices. Second we filled a glass capillary of known thickness (10\,\textmu m $\pm 10\%$) with a solution of fluorescent beads ($\O\,0.5\,$\textmu m, Latex beads, Sigma) and measured the geometrical path length, by the use of a calibrated piezo element, calculating the correction factor $n_{k}$ based on the difference of the capillary and geometrical length:
	$n_k=\frac{n_{buffer}}{n_{oil}}\approx\frac{h_{capillary}}{h_{geometrical}}=0.9 \pm 0.04$. Both methods gave equivalent results.
Data were taken 30-80\,\textmu m into the sample to reduce wall effects.

Our shear protocol consists of four blocks of five minutes where a shear rate of 0.075, 0.15, 0.225, 0.3\,s$^{-1}$, respectively, is applied. The reason for the shear protocol is that after sample loading and inserting the cone, the orientation of the filaments for $c_{high}$ is not well defined, while it does not relax. This is a fact that cannot be avoided. Thus quite some strain units are needed to remove this memory effect. For $c_{low}$ the sample does relax before we start the experiment and indeed we can follow trends with increasing strain.

Rheology data are taken with an Anton Paar MCR501, using a cone-plate geometry with 30 mm diameter and 1 degree angle.

\subsection{Data analysis}
The 3D filament contours were tracked with the autodepth-function of the visualization and analysis software Imaris (Bitplane/ Imaris 6.1), and subsequent filament position analysis was done using Matlab (Version, Mathworks).
As a control for the conformation of the filaments before the shear experiment ($\gamma=0$) one frame was taken before placing the cone. This frame contained about 50 tracer filaments, that were tracked with Imaris.
For the analysis of the filament conformation at a certain strain value, 3 statistically independent stacks (circa 30\,s apart) before changing the shear rate, were analyzed, containing also about 50 tracer filaments each, corresponding to a strain of $\gamma$=21$\pm$ 3, 59 $\pm$ 6, 126 $\pm$ 7 and 215 $\pm$ 12.
For the last data point 3 frames, 120\,s, 150\,s and 180\,s after cessation, were analyzed.

\section{Acknowledgments}
G.H.K. was supported by the Foundation for Fundamental Research on Matter (FOM), which is part of NWO, and N.A.K. acknowledges support by a Marie Curie FP7 IIF Fellowship. D.G. was supported by  the  International  Helmholtz Research School (BioSoft). We thank F. MacKintosh, R. Winkler, J. Dhont, M. Giesen and R. Merkel for fruitful discussions and Y. Jia for help with the data analysis.



\end{document}